\documentclass[fleqn,11pt]{wlscirep}
\usepackage{setspace}
\title{Discovery of a very Lyman-$\alpha$-luminous quasar at z=6.62}

\author[1,*]{Ekaterina Koptelova}
\author[1,*]{Chorng-Yuan Hwang}
\author[1,+]{Po-Chieh Yu}
\author[1,+]{Wen-Ping Chen}
\author[1,+]{Jhen-Kuei
Guo}\affil[1]{National Central University, Graduate Institute of
Astronomy, Taoyuan City, 32001, Taiwan}

\affil[*]{koptelova@astro.ncu.edu.tw}
\affil[*]{hwangcy@astro.ncu.edu.tw} \affil[+]{these authors
contributed equally to this work}

\begin{abstract}
Distant luminous quasars provide important information on the
growth of the first supermassive black holes, their host galaxies
and the epoch of reionization. The identification of quasars is
usually performed through detection of their Lyman-$\alpha$ line
redshifted to $\gtrsim$0.9 microns at z$>$6.5. Here, we report the
discovery of a very Lyman-$\alpha$ luminous quasar, PSO
J006.1240+39.2219 at redshift z=6.618, selected based on its red
colour and multi-epoch detection of the Lyman-$\alpha$ emission in
a single near-infrared band. The Lyman-$\alpha$ line luminosity of
PSO J006.1240+39.2219 is unusually high and estimated to be
0.8$\times$$10^{12}$ Solar luminosities (about 3\% of the total
quasar luminosity). The Lyman-$\alpha$ emission of PSO
J006.1240+39.2219 shows fast variability on timescales of days in
the quasar rest frame, which has never been detected in any of the
known high-redshift quasars. The high luminosity of the
Lyman-$\alpha$ line, its narrow width and fast variability
resemble properties of local Narrow-Line Seyfert 1 galaxies which
suggests that the quasar is likely at the active phase of the
black hole growth accreting close or even beyond the Eddington
limit.
\end{abstract}
\begin{document}

\flushbottom
\maketitle
%
%
\thispagestyle{empty}


\section*{Introduction}

High-redshift quasars provide important constraints on theories of
structure formation and properties of the intergalactic medium
(IGM) at early cosmic epochs. The most distant of them are seen at
redshifts z$>$6.5 (when the Universe was less than 6\% of its
present age), before the end of reionization\cite{Becker2001}.
There are only seven quasars found at z$>$6.5. Four were
discovered in near-infrared surveys\cite{Mortlock,Venemans2013}.
Three new quasars\cite{Venemans2015} were discovered recently from
the 3$\pi$ Panoramic Survey Telescope and Rapid Response System
\cite{Kaiser2010,Schlafly2012} (Pan-STARRS1 or PS1) sensitive up
to the near-infrared y$_{\rm PS1}$ band ($\lambda_{\rm
eff}$=0.96$\mu$m). These highest-redshift quasars host the most
massive supermassive black holes (SMBHs) with masses
10$^{9}$-10$^{10}$ M$_{\bigodot}$ (where M$_{\bigodot}$ is the
solar mass) accreting close to the Eddington
luminosity\cite{Mortlock,Wu2015,DeRosa}.

We used the PS1 data to search for new quasars at z$>$6.5 among
extremely red sources with colours z$_{\rm PS1}$ - y$_{\rm PS1}$
$>$ 2~mag and multiple detections at several different epochs (see
'Quasar candidate selection' in the Methods). In our follow-up
observations of these sources on November 2, 2015 we discovered a
new high-redshift quasar, PSO J006.1240+39.2219, at z=6.618 (see
'Spectroscopic follow-up' and 'Redshift measurements' in the
Methods).

\section*{Results}

The discovery spectrum of PSO J006.1240+39.2219 shows an unusually
strong, compared to the continuum and, at the same time, narrow
Lyman-$\alpha$ (Ly$\alpha$) emission line with a line width of
1300$\pm$90 km s$^{-1}$ (see Fig. 1). This line width is smaller
than the typical width of the emission lines produced in the Broad
Line Region (BLR) of quasars. For comparison, we estimated the
Ly$\alpha$ line width of the quasars known at z$>$6.5 to be
between 2000 and 3800 km s$^{-1}$. The observed metal lines of PSO
J006.1240+39.2219 (NV, OI+SiII and CII) do not exhibit the unusual
strength as its Ly$\alpha$ line. As seen in Fig.~1, the peak flux
ratio of the Ly$\alpha$ and NV lines of PSO J006.1240+39.2219 is
more than twice larger than that observed in high-redshift quasars
with strong Ly$\alpha$ emission\cite{banados2016}. The small width
of the Ly$\alpha$ line of PSO J006.1240+39.2219 is difficult to
explain as due to particularly strong absorption by neutral
hydrogen (HI). The absorption features on the blue side of the
line have a minor impact on its strength and shape (the core of
the line is rather symmetric relative to the redshifted line
wavelength). This might imply that the Ly$\alpha$ line of PSO
J006.1240+39.2219 is dominated by a narrow-line component and its
profile is intrinsically narrow. In the local Universe, narrow
high-ionization, broad emission lines are also observed in
Narrow-Line Seyfert 1 (NLS1) galaxies. As demonstrated in Fig.~1,
the Ly$\alpha$ line of the NLS1s is much stronger than in
broad-line quasars\cite{Puchnarewicz1998}. The NLS1 galaxies have
smaller black holes of 10$^{6}$ -10$^{8}$ M$_{\bigodot}$ resulting
in the narrow width of the BLR lines. Most of the NLS1s accrete at
the super-Eddington limit\cite{Mathur2000,Boroson2002,Bian2004}.

From power-law fit F$_{\lambda}\sim \lambda^{\alpha}$ to the
continuum of PSO J006.1240+39.2219 between 9500-9900 and
10000-10150 \AA, we estimated a spectral slope of
$\alpha$=-1.10$\pm$0.48. By extrapolating the power law to
1450$\times$(1 + z)\AA, we measured the absolute magnitude of the
quasar at rest-frame wavelength 1450\AA $  $  to be M$_{1450}$ =
-26.1$\pm$0.4, where the error includes the uncertainty in the
spectral slope and redshift. Applying a bolometric correction
factor of 4.4 to the ultraviolet (UV)
luminosity\cite{Richards2006}, we estimated a total quasar
luminosity of 2.8 $\times$ 10$^{13}$ L$_{\bigodot}$ (where
L$_{\bigodot}$ is the solar luminosity). The relation between
black hole mass and bolometric luminosity for the known z$\sim$6
quasars follows well the expected relation for accretion at the
Eddington limit\cite{Willott2010}. If PSO J006.1240+39.2219
accretes at the Eddington limit, its luminosity implies a black
hole mass of 10$^{8}$--10$^{9}$
M$_{\bigodot}$\cite{Wu2015,Willott2010}. Given the relation
between mass of black holes and square of the width of broad
emission lines\cite{Peterson1999}, the narrow Ly$\alpha$ line of
PSO J006.1240+39.2219 implies up to an order of a magnitude
smaller black hole mass than expected from the quasar luminosity,
and the super-Eddington accretion rate\cite{Wu2015,Willott2010}.

From the spectrum of PSO J006.1240+39.2219 with the subtracted
continuum we measured its Ly$\alpha$ line luminosity to be 0.8
$\times$ 10$^{12}$ L$_{\bigodot}$, which is about 3\% of the total
luminosity of the quasar. We compared the luminosity of the
Ly$\alpha$ line of PSO J006.1240+39.2219 with that of the other
quasars discovered at z$>$6.5. As shown in Fig.~2, the Ly$\alpha$
luminosity of PSO J006.1240+39.2219 is larger than the Ly$\alpha$
luminosity of the z$>$6.5 quasars by more than a factor of two. It
is also more than ten times larger than the Ly$\alpha$ luminosity
of the most luminous Lyman Alpha Emitting galaxies (LAEs) seen
during the epoch of reionization\cite{Hu2016}. The relative
contribution of the Ly$\alpha$ line into the total quasar
luminosity is somewhat uncertain as due to the uncertainty in the
continuum fit. However, we calculated that for power-law slopes
between -0.5 and -2.5, the Ly$\alpha$ emission always dominates
the UV continuum and contributes 2--4\% into the bolometric
luminosity of the quasar. In the other z$>$6.5 quasars this
contribution is only 0.1-0.5\% (see Fig.~3). The rest-frame
equivalent width (EW) of the Ly$\alpha$ line of PSO
J006.1240+39.2219 is also large. Similar to the previous studies,
we estimated the EW of the quasar by integrating the line flux
above the continuum within 1160$<\lambda_{\rm rest}<$1290\AA,
which includes the Ly$\alpha$ and NV lines. The measured EW of PSO
J006.1240+39.2219 is equal to 182\AA $ $ and corresponds to the
high end of the equivalent width distribution of the known z$>$5.6
quasars with the peak at EW$\approx$35\AA\cite{banados2016}. Many
of the z$>$5.6 quasars have absolute magnitudes brighter than that
of PSO J006.1240+39.2219, but only small fraction of them exhibit
strong Ly$\alpha$ emission lines (Their composite spectrum is
shown in Fig.~1). The typical EW of these quasars is
$\sim$140\AA\cite{banados2016}. The EWs of the known z$>$6.5
quasars are smaller than 35\AA (except for PSO J338+29 with
EW$\simeq$70\AA\cite{Venemans2015}), i.e., corresponds to the
lower end of the EW distribution of the high-redshift quasars,
which can be explained by stronger HI absorption at z$>$6.5. The
large EW of PSO J006.1240+39.2219 compared to these quasars
implies weaker nearby intergalactic HI absorption.

The multi-epoch photometry of PSO J006.1240+39.2219 shows the
change of the quasar brightness. From the PS1 images taken between
June 2010 and July 2013, we measured the y$_{\rm PS1}$-band
brightness of the quasar at different epochs.  The resulting
quasar light curve is shown in Fig.~4. The brightness of the
quasar at the time of our spectroscopic observations is also
presented. From the quasar light curve we find that the quasar is
variable on rest-frame timescales of days and months, with an
amplitude exceeding its multi-epoch mean brightness by more than
2.5$\sigma$. The overall peak-to-peak amplitude of the observed
variations is $\sim$0.7 mag. Between 2010 and 2011, the quasar
became brighter by about 0.24 mag within 50 days in the quasar
rest frame. In 2013, PSO J006.1240+39.2219 changed its brightness
from 20.15$\pm$0.09 to 19.66$\pm$0.07 mag over a period of
$\sim$2~days in the quasar rest frame. These high-amplitude
variations are larger than brightness changes of 0.1-0.2~mag
expected from the UV/optical structure function\cite{Ivezic2004}
and damped random walk model\cite{MacLeod2012} of quasar
variability on similar timescales. However, we note, that the
Ly$\alpha$ line is ionized by the extreme UV and soft X-ray
radiation which can be highly variable. For instance, the soft
X-ray flux of NLS1s can change by a factor of ten on timescales of
days\cite{Turner2001,Romano2002}. The variation in Ly$\alpha$
emission can occur almost simultaneously with the variation of the
ionizing flux on short timescales limited by hydrogen
recombination time $\tau_{\rm rec}=(n_{\rm e} \alpha_{\rm
B})^{-1}\approx 40(n_{\rm e}/10^{11} \rm cm^{-3})^{-1}$~s (where
the typical electron density of the broad-line emitting gas is
$n_{\rm e}$$\gtrsim$10$^{8}$ cm$^{-3}$\cite{Peterson}) and the
size of the Ly$\alpha$ emitting
region\cite{Peterson2000,Collier2001}.

The y$_{\rm PS1}$ band measures the total flux from the Ly$\alpha$
line and nearby UV continuum. Therefore, observed variability of
PSO J006.1240+39.2219 can be caused both by the line and continuum
variations. However, the continuum brightness, corresponding to
the y$_{\rm PS1}$-band multi-epoch mean quasar flux, is
$\gtrsim$21 mag, which is below the detection limit for single
exposures in the 3$\pi$ PS1 survey (m$_{y_{\rm PS1}}^{\rm
lim}$=20.12~mag )\cite{Chambers2006}. The relative flux
contribution of the Ly$\alpha$ line into the y$_{\rm PS1}$-band
total flux of the quasar is more than 70\%.  Therefore, the
observed flux mostly comes from the Ly$\alpha$ line of PSO
J006.1240+39.2219. The rapid y$_{\rm PS1}$-band variations of the
quasar provide the evidence of variable Ly$\alpha$ emission which
responds fast to the variations of the extreme UV and soft X-ray
flux and, therefore, originates close to the central
engine\cite{Gaskell}. The small size of the Ly$\alpha$ emitting
region, as expected from variability of the quasar, suggests a
rather small mass of the central black hole\cite{Peterson1999}.

\section*{Discussion}

We reported the discovery of the Ly$\alpha$-luminous narrow-line
quasar, PSO J006.1240+39.2219, with the first evidence of
broad-band quasar variability at high redshift. We find a
similarity between the properties of PSO J006.1240+39.2219 and the
NLS1 galaxies. The NLS1s exhibit rapid UV variability and narrow
broad lines, as a result of the smaller black hole masses, an
order of a magnitude smaller than the black hole masses of the
broad-line quasars of the same luminosities. Similar to the NLS1s,
the strong narrow Ly$\alpha$ line of PSO J006.1240+39.2219 without
a prominent broad-line component and its short-term variability
provide the evidence of the smaller black hole mass of this quasar
than that expected from the luminosity - black hole mass relation.

The high luminosity of the Ly$\alpha$ line of PSO
J006.1240+39.2219 implies that the extreme UV and soft X-ray
component of the quasar continuum is strong and sustains its
Ly$\alpha$ emission at a very high level\cite{Krolik,Green}. We
estimate the luminosity of this high-energy continuum to be
L$_{\rm ion}$ = 1.8 L(Ly$\alpha$)/f$_{\rm esc}^{\rm
Ly\alpha}\approx$5$\times$10$^{12}$L$_{\bigodot}$\cite{Orsi},
where we assume that the average energy of ionizing photons is
13.6eV and escape fraction of the Ly$\alpha$ photons is f$_{\rm
esc}^{\rm Ly\alpha}\approx$0.3. The adopted escape fraction
represents the volume-averaged value that is found to evolve
approximately as power law f$_{\rm esc}^{\rm
Ly\alpha}\propto(1+z)^{2.57}$ between redshift 0 and
6\cite{Hayes2011}. We note, that the volume-averaged escape
fraction includes effects of absorption by the IGM that might lead
to the smaller values of f$_{\rm esc}^{\rm Ly\alpha}$ at z$>$6. In
spite of the uncertainty in f$_{\rm esc}^{\rm Ly\alpha}$, being
the most luminous Ly$\alpha$ emitter, PSO J006.1240+39.2219 is the
powerful source of ionizing radiation which likely  has an
important contribution into ionization of the IGM surrounding the
quasar. From the observed spectrum we measure the size of the
quasar ionized HII region scaled to M$_{1450}$ = -27 to be R$_{\rm
NZ}$=4$\pm$1 Mpc, which is slightly larger than a near zone of
2.5-3.5 Mpc expected from the empirical relation between R$_{\rm
NZ}$ and redshift\cite{Carilli} (see 'The near-zone size' in the
Methods).

The observed y$_{\rm PS1}$-band brightness variations of PSO
J006.1240+39.2219 are likely due to variability of its Ly$\alpha$
emission as it substantially dominates the y$_{\rm PS1}$-band flux
of the quasar. The size of the Ly$\alpha$ emitting region of PSO
J006.1240+39.2219 inferred from the timescale of the Ly$\alpha$
rapid variations is about 2 light days. This is similar but
slightly smaller than the BLR regions of the local NLS1
galaxies\cite{Collier2001,Grier2012,Rafter2013}. From the
observations of reverberation time lags between the UV/X-ray
continuum and Ly$\alpha$ line (and also between the UV/X-ray
continuum and Balmer lines), the typical size of the Ly$\alpha$
emitting region of the NLS1s is estimated to be 3-10 light days.
For comparison, the time lags (and correspondingly the BLR sizes)
in broad-line quasars are $\geqslant$ 1 month
\cite{Kaspi,Peterson2004}. We caution that if the UV continuum of
PSO J006.1240+39.2219 is highly variable (e.g., changing by about
1~mag on short timescales) its variability imposed on the
variations of the Ly$\alpha$ flux would lead to underestimation of
the size of the Ly$\alpha$ emitting region inferred from the
observed short-term variations.

From the similarity of PSO J006.1240+39.2219 with the NLS1
galaxies we infer that this quasar is young, at the early phase of
its black hole and bulge formation. These Ly$\alpha$-line luminous
young quasars seen at early cosmic epochs might be capable of
ionizing large volumes of gas and might play a significant role in
cosmic reionization.

\section*{Methods}

\noindent \textbf{Quasar candidate selection.} We searched for
z$_{\rm ps1}$-band dropouts in the first and second internal data
releases of the PS1 survey (PV1 and PV2) using the i$_{\rm ps1}$-,
z$_{\rm ps1}$- and y$_{\rm ps1}$-band photometric catalogues.
First, from the y$_{\rm ps1}$-band catalogue we selected point
sources assuming that the difference between their point spread
function (PSF) and aperture magnitudes is less than 0.3 mag, and
the chi-square of the PSF fit is $\chi_{\rm red}^{2}$ $<$ 1.5.
From the resulting sample we selected the z$_{\rm ps1}$-band
dropout quasar candidates using the following criteria: z$_{\rm
ps1}$-y$_{\rm ps1}$ $>$ 2, $\sigma_{\rm y_{\rm PS1}}$ $<$ 0.1 mag
(where $\sigma_{\rm y_{\rm PS1}}$ is the y$_{\rm ps1}$-band
photometric error), i$_{\rm ps1}$ $>$24 and z$_{\rm ps1}$ $>$24
mag. These criteria are similar to those adopted in the previous
searches of high-redshift quasars from PS1\cite{Venemans2015}.
Unlike the previous works, we additionally checked for multi-epoch
detections of our z$_{\rm ps1}$-band dropout candidates in the
y$_{\rm ps1}$-band. The PS1 survey conducted repeated scans of the
sky and provided multi-epoch photometry for detected sources in
all PS1 bands. The z$_{\rm ps1}$-dropout candidates detected at
least at two different epochs were considered by us as reliable.
In this way we excluded short-lived transients and other possible
artifacts from our colour-selected sample. The strongest of the
multi-epoch candidates had photometric measurements at five
different epochs while no detection in the z$_{\rm PS1}$ band.
This candidate is the high-redshift quasar presented in this work.
The selected high-redshift quasar candidates were also checked for
the counterparts in the Wide-Field Infrared Survey Explorer
all-sky source catalogue\cite{Wright} (AllWISE) within a match
radius of 3 arcsec. However, none of them was detected in the WISE
bands. Using this result, we place upper limits on their WISE W1
and W2 magnitudes to be W1$>$19.7 and W2$>19.3$~mag (i.e., fainter
than the WISE W1 and W2 limiting magnitudes).

\noindent \textbf{Spectroscopic follow-up.}  We performed
simultaneous photometric and spectroscopic observations of twelve
z$>$6.5 quasar candidates with the Subaru Faint Object Camera And
Spectrograph\cite{Kashikawa2002} (FOCAS) of the 8.2-m Subaru
telescope. The observations were carried out on November 2, 2015.
We used FOCAS long-slit mode, VPH900 grating and the SO58 order
cut filter, giving us a wavelength coverage of 7500-10450\AA  $ $
and a dispersion of 0.74 \AA $ $ pixel$^{-1}$. The slit was 0.8
arcsec wide resulting in a spectroscopic resolution of
R$\sim$1500. The seeing during the observations varied between
0.26-0.48 arcsec. Prior to spectroscopy we took acquisition images
of the candidates in the FOCAS Y band. The spectra were taken only
for three candidates, two of which were reliably detected during
acquisition. Out of these three targets, only one had the blue-end
cutoff typical for high-redshift sources. We took five 1000s
through-slit exposures of this target which was identified as a
quasar based on its spectrum. The stacked spectrum of the quasar
and its uncertainty were calculated using median combine of the
individual exposures. The quasar spectrum was absolute flux
calibrated using the spectrophotometric standard star BD+28d4211
observed on the same night.

\noindent \textbf{Redshift measurements.} To estimate the redshift
of the quasar, we measured the redshifted positions of the NV,
OI+SiII and CII emission lines at $\lambda_{\rm rest}$=1239.85,
1305.42 and 1336.60 \AA. The redshift was determined by
calculating the cross-correlation function between region
9300-10200\AA $ $ of the quasar spectrum and the redshifted
composite SDSS quasar spectrum\cite{vandenberk}. The fitted
wavelength range did not include the Ly$\alpha$ line. The best
correlation with a correlation coefficient of 0.86 was achieved
for a redshift of z=6.618$\pm$0.02.

\noindent \textbf{The near-zone size.} To measure the size of the
ionized HII region around the quasar, we first smoothed the quasar
spectrum to a resolution of 20\AA. The transmission in the near
zone was calculated by dividing the smoothed spectrum by power-law
continuum   F$_{\lambda}\sim \lambda^{-1.1}$, and the Lorenzian
and Gaussian fitted to the Ly$\alpha$ and NV lines. When measuring
the near zone, we corrected the quasar redshift for a systematic
offset of about +0.02. This systematic offset was reported between
the MgII redshift and redshifts derived from high-ionization
lines\cite{richards2002}. We measured a proper near zone of
3$\pm$1 Mpc as a region where the transmitted flux drops below
10\% of extrapolated continuum emission\cite{fan2006}. The size of
the near zone scaled to M$_{1450}$ = -27 is 4$\pm$1
Mpc\cite{Carilli}.



\section*{Acknowledgements}

The research project was supported by the Ministry of Science and
Technology of Taiwan, grant No. MOST 103-2119-M-008-017-MY3 and
MOST 105-2811-M-008-074. The Pan-STARRS1 Surveys have been made
possible through contributions of the Institute for Astronomy,
University of Hawaii, the Pan-STARRS Project Office, the
Max-Planck Society and its participating institutes,
Max-Planck-Institute for Astronomy, Heidelberg and
Max-Planck-Institute for Extraterrestrial Physics, Garching, The
Johns Hopkins University, Durham University, University of
Edinburgh, Queens University Belfast, Harvard-Smithsonian Center
for Astrophysics, the Las Cumbres Observatory Global Telescope
Network Incorporated, the National Central University of Taiwan,
the Space Telescope Science Institute, the National Aeronautics
and Space Administration under grant No. NNX08AR22G issued through
the Planetary Science Division of the NASA Science Mission
Directorate, the National Science Foundation under grant
AST-1238877, the University of Maryland, and Eotvos Lorand
University (ELTE). This research has made use of the services of
the ESO Science Archive Facility.

\section*{Author contributions statement}

E.K. selected the source, observed and analysed the data. C.-Y. H.
supervised the data interpretation and edited the manuscript.
P.-C. Y. contributed to the calculations and edited the
manuscript. W.-P. C. and J.-K.G. gave data support for conducting
the source selection. All authors discussed the results and
implications and commented on the manuscript at all stages.

\section*{Additional information}

\textbf{Competing financial interests} The authors declare that
they have no competing financial interests.

\begin{figure}[ht]
\centering
\includegraphics[width=\linewidth]{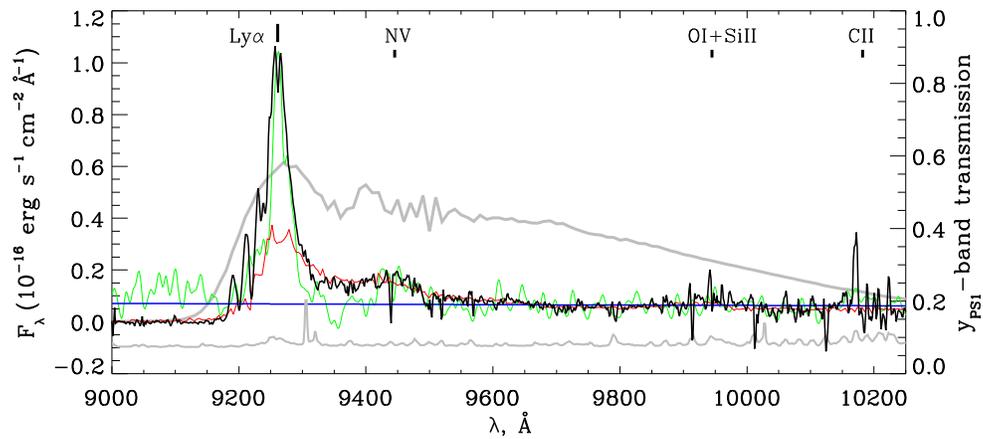}
\caption{FOCAS spectrum of PSO J006.1240+39.2219 (black line). The
displayed spectrum is smoothed with the Gaussian filter using a
smoothing window of 5\AA.  The sigma spectrum shown with a grey
line is offset for better representation. The Ly$\alpha$ line is
detected with a signal-to-noise ratio (SNR) of 34. The SNR ratios
of the spectrum at the positions of the NV, OI+SiII and CII
emission lines are 15, 4 and ~1. The widths of the Ly$\alpha$
line, deblended and fitted with the Lorenzian profile, is
estimated to be 1300 km s$^{-1}$. The composite spectrum of
z$>$5.6 quasars with strong Ly$\alpha$ emission\cite{banados2016}
is overplotted in red. The redshifted UV spectrum of the NLS1
galaxy RE J1034+396 at z=0.043 is shown in green. (The spectra are
scaled to the NV emission line of PSO J006.1240+39.2219). The
power-law continuum fit over spectral windows 9550 - 9900 and
10000 - 10150 \AA $ $ (F$_{\lambda}\sim \lambda^{-1.1}$) is shown
with a blue line. The transmission curve of the PS1 y-band filter
is plotted with a thick grey line.} \label{fig:spectrum}
\end{figure}

\begin{figure}[ht]
\centering
\includegraphics[width=\linewidth, scale=0.4]{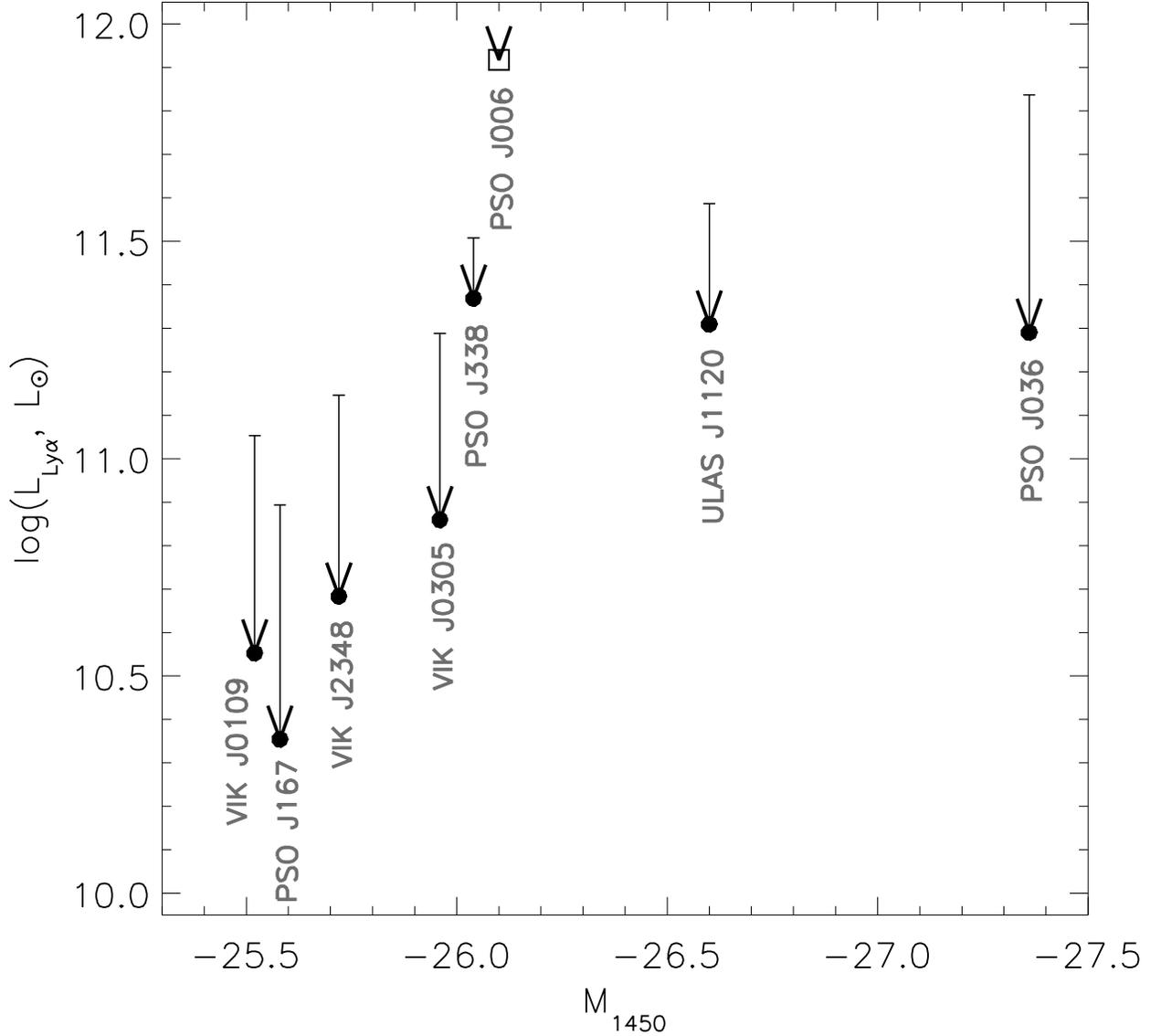}
\caption{Absolute UV magnitude M$_{1450}$ of quasars at redshidft
z$>$6.5 versus Ly$\alpha$ line luminosity. The quasars ULAS
J1120+0641\cite{Mortlock}, VIK J0109-3047\cite{Venemans2013}, VIK
J2348-3054\cite{Venemans2013}, VIK J0305-3150\cite{Venemans2013},
PSO J167-13\cite{Venemans2015}, PSO J338+29\cite{Venemans2015},
PSO J036+03\cite{Venemans2015} are marked with black circles, PSO
J006.1240+39.2219 is shown with an open square. The Ly$\alpha$
line luminosity is estimated by integrating the line flux between
1204--1229\AA ($\sim$ 6170 km s$^{-1}$)\cite{woo2013}. The upper
limits correspond to the luminosity of the Ly$\alpha$ line
obtained without continuum subtraction from the line flux. The
continuum contribution into the Ly$\alpha$ line of PSO
J006.1240+39.2219 is estimated to be less than 10\% (within the
region marked by the square) and is significantly less than in the
other z$>$6.5 quasars. } \label{fig:lum}
\end{figure}

\begin{figure}[ht]
\centering
\includegraphics[width=\linewidth, scale=0.4]{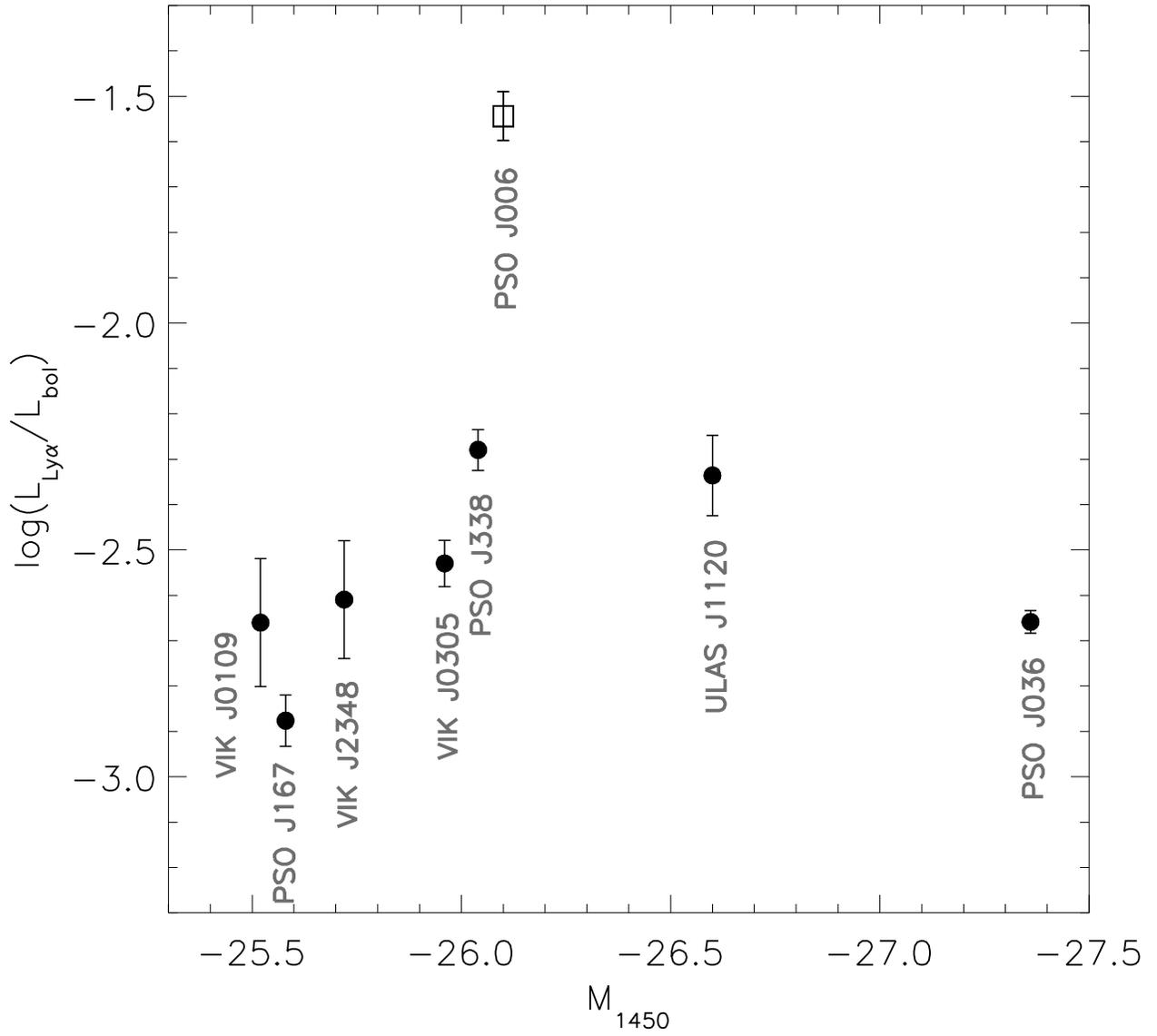}
\caption{Absolute UV magnitude M$_{1450}$ of quasars at redshidft
z$>$6.5 versus Ly$\alpha$ line luminosity expressed as a fraction
of the bolometric luminosity. The object notation is similar to
the previous figure. } \label{fig:slope}
\end{figure}

\begin{figure}[ht]
\centering
\includegraphics[width=\linewidth]{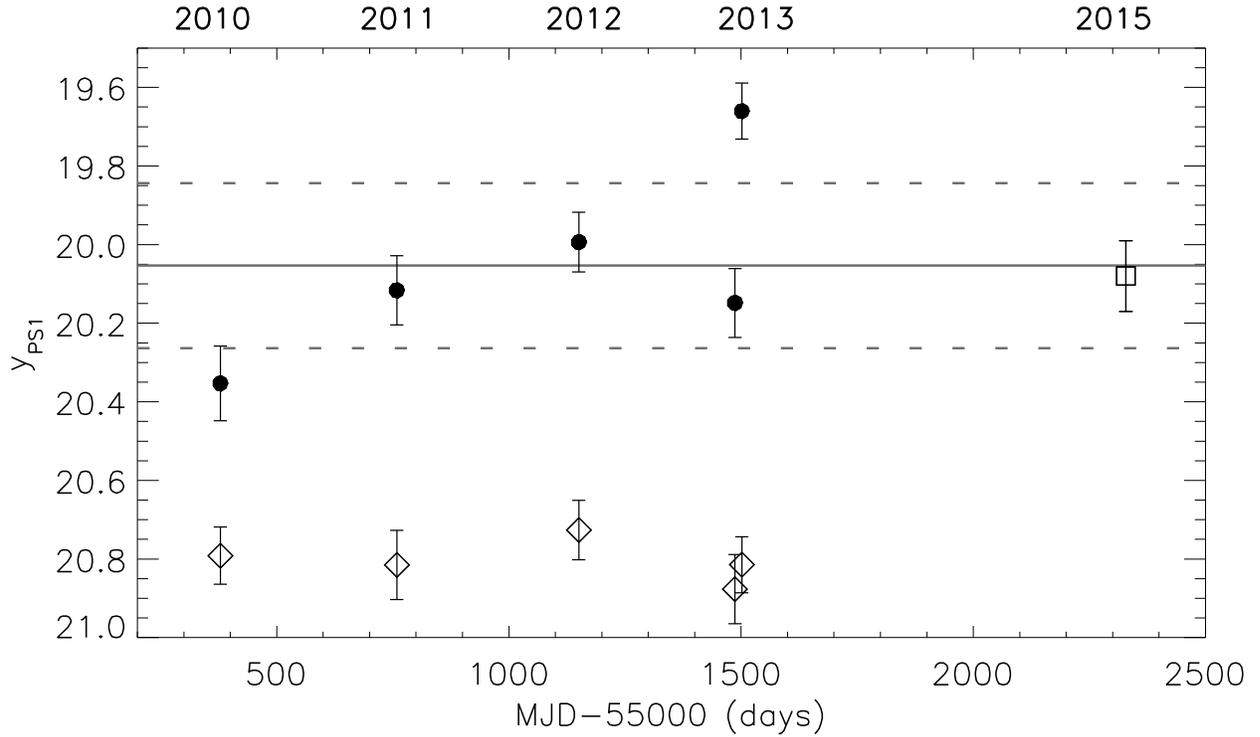}
\caption{The y$_{\rm PS1}$-band light curve of PSO
J006.1240+39.2219. Black points show the quasar brightness
measured using aperture photometry of the PS1 images taken between
June 2010 and July 2013. At each of the epochs, the quasar was
observed at least two times for a total of 60s. The measured
quasar flux and uncertainty at the different epochs are estimated
as the mean and standard deviation of five independent quasar flux
measurements relative to five nearby faint stars of $\sim$
19.5--20.0 mag. The solid and dotted grey lines correspond to a
multi-epoch mean and its $\pm$2.5$\sigma$ error of 20.05$\pm$0.21
mag. In 2010 the quasar was fainter of its mean brightness by 0.24
mag. Within 15 days in 2013, it became brighter than its mean
brightness by $\sim$0.4 mag (which is 4.6$\sigma$ off the
multi-epoch mean), showing short-term variability. The overall
brightness change between 2010 and 2013 is $\sim$0.7 mag. The open
square shows the brightness of the quasar in 2015 estimated from
its discovery spectrum by integrating the flux through the y$_{\rm
PS1}$ passband. The light curve of one of the nearby faint stars
of 19.65$\pm$0.07 mag chosen for quasar flux calibration, is shown
with diamonds. (For better representation the light curve of the
star is shifted by +1.15 mag.) } \label{fig:lc}
\end{figure}

\end{document}